\documentstyle[12pt,a4]{article}
\input{epsf}
\newcommand{\plus}{\makebox[15pt][c]{$+$}}
\newcommand{\minus}{\makebox[15pt][c]{$-$}}
\newcommand{\errr}[2]{\raisebox{0.08em}{\scriptsize
{$\;\begin{array}{@{}l@{}}
                          \plus\makebox[0.9em][r]{#1} \\[-0.12em]
                          \minus\makebox[0.9em][r]{#2}
                        \end{array}$}}}
\newcommand{\err}[2]{\raisebox{0.08em}{\scriptsize
{$\;\begin{array}{@{}l@{}}
                          \plus\makebox[0.55em][r]{#1} \\[-0.12em]
                          \minus\makebox[0.55em][r]{#2}
                        \end{array}$}}}
\newcommand{\ewxy}[2]{\setlength{\epsfxsize}{#2}\epsfbox[10 60 640
570]{#1}}
\newcommand{\beq}{\begin{equation}}
\newcommand{\eeq}{\end{equation}}

\begin{document}
\begin{titlepage}

\begin{flushright}
{\small GUTPA/95/12/1\\
FSU-SCRI-95-121\\
OHSTPY-HEP-T-95-026\\
December 1995}
\end{flushright}

\vspace*{2mm}

\begin{center}
{\Huge The Heavy-Light Spectrum from\\
 Lattice NRQCD }\\[9mm]

{ \large  A.~Ali Khan~\footnote{UKQCD Collaboration} and
C.~T.~H.~Davies~${\vphantom{\Large A}}^1$} \\
University of Glasgow, \\
Glasgow G12~8QQ, UK \\
\vspace{3mm}
{\large  S. Collins~\footnote{Present address: The University of
Edinburgh,
Edinburgh EH9~3JZ, U.K.} and J. Sloan} \\
Supercomputer Computations Research Institute \\
Florida State University \\
Tallahassee, FL 32306-4052, USA \\
\vspace{3mm}
{\large  J. Shigemitsu} \\
The Ohio State University \\
Columbus, OH 43210, USA \\
\end{center}
\vspace{2mm}
\begin{abstract}
{We present a lattice investigation of heavy-light mesons in the
quenched
approximation,
 using non-relativistic
QCD for the heavy quark and a clover improved Wilson
formulation for the light quark. A comprehensive calculation of the
heavy-light
spectrum has been performed for various heavy quark masses around the
$b$. Our results for
the $B_s-B_d$ splitting agree well with the
experimental value. We find the $\Lambda_b-B$ splitting to be
compatible
 with experiment, albeit with large error bars. Our  $B^*-B$
splitting is
slightly low, which could be explained as an effect of quenching. For
the first
time, we are able to estimate the mass of $P$ states at the $B$ and
compare them
 with experiment. }
\end{abstract}
\vspace{3cm}
PACS numbers: 12.38.Gc, 14.65.Fy, 14.40.Nd, 14.20.Mr
\thispagestyle{empty}
\end{titlepage}
\newpage
\section{\label{sec:intro} Introduction}
Heavy-light hadrons containing a $b$ quark are of topical
interest in particle theory. Theoretical predictions for heavy-light
meson
 decay constants are needed for an experimental determination of
elements of
the CKM matrix~\cite{CKM}. A
theoretically interesting feature of heavy-light hadrons is that, in
the
limit of a very heavy quark mass $m_Q$, the dynamics simplify and can
be described by
an effective field theory (Heavy Quark Effective Theory)~\cite{HQET}.
In the
 heavy quark limit $m_Q \rightarrow \infty$ additional symmetries
between
states of different spin and flavour
appear which make it possible to relate different decay processes to
each
other and to make qualitative predictions about the spectrum. Away
from the
heavy quark limit a systematic expansion in
the heavy quark mass can be performed whose coefficients have to be
determined non-perturbatively. On the lattice this can be done from
first
principles.

Since, for the values of the lattice spacings which can be
 achieved in present computer simulations, the $b$ quark mass is $>
1$ in lattice units, $b$ quarks cannot be simulated directly using a
na\"{\i}ve relativistic quark action. It has been suggested that
heavy quarks
around and above the charm can be simulated using a re-formulation of
a
relativistic quark action~\cite{mackenzie}, but there are indications
that
there are problems with this method in the $b$ quark
region~\cite{sara_lat95},
at least in the way the method has been implemented to date.
So simulations are often  done at smaller
quark masses, up to the charm, and results extrapolated to the $b$.
With this
method it is very difficult to get certain spectral quantities,
such as spin splittings, to agree with experiment. Simulations of
heavy-light
systems in the static (i.e. infinite mass) limit of the heavy quark
give
sensible results only  for spectral quantities that are fairly
independent of $m_Q$. For other quantities, contributions due to the
finiteness
of the heavy quark mass have to be calculated directly as
perturbations in
$1/m_Q$ (see e.g.~\cite{bbar}). In addition the signal-to-noise ratio
is much worse than for propagating quarks.

The non-relativistic formulation of QCD (NRQCD)~\cite{Lepage92}
however allows
us, in principle,
to simulate at the $b$ quark mass.
The largest momentum scale which governs discretization
errors is not  the heavy quark mass but the non-relativistic momentum
of the
heavy
quark, which is much smaller. This method has been successful in
precision
spectroscopy of heavy-heavy systems such as $\Upsilon$ and
$J/\Psi$ mesons~\cite{upsilon,jpsi}. In these
systems the energy scale inside the meson is set by the kinetic
energy of the
heavy quarks, which is of the order of $\Lambda_{QCD}$. Relativistic
corrections are included in a systematic expansion in $v_Q^2$, where
$v_Q$ is the
velocity of the heavy quark. For mesons with one heavy and one light
quark
the power counting rules for the non-relativistic expansion are
determined by
the fact that here the energy scale for the heavy quark inside the
meson is
the binding energy. For $m_Q \rightarrow \infty$ only the time
derivative
appears in the tree level Langrangian. The covariant
time derivative acting on the
heavy quark spinor gives the quark energy:
\beq
D_t Q\sim E_{\rm bind}Q,
\eeq
Due to momentum conservation the heavy and light quark momenta are
equal in the
rest frame of the meson:
\beq
m_Q v = p_Q = p_q \sim O(\Lambda_{QCD}),
\eeq
 where  $\Lambda_{QCD}$ takes a value
around $300-500$ MeV. So the heavy quark velocity is small:
\beq
v_Q \sim O\left(\frac{\Lambda_{QCD}}{m_Q}\right) ,
\eeq
which is of the order of $10 \%$ for the $B$,
and we can choose $v_Q$, or alternatively $1/m_Q$, as the
expansion parameter for
the NRQCD interactions. Now we have to determine the  order in $v$ of
various
correction terms. We can immediately write down the
operators which appear with a tree level $1/m_Q$ coefficient: the
non-relativistic  kinetic energy operator
\beq
O_{\rm kin} = -\frac{\vec{D}^2}{2m_Q}
\eeq
and the magnetic interaction operator
\beq
O_{\rm mag} =  \frac{\vec{\sigma}\cdot\vec{B}}{2m_Q}.
\eeq
To estimate the size of the heavy quark kinetic energy, we write
\beq
\frac{\vec{p}^2}{2m_Q} = \frac{(m_Qv)^2}{2m_Q}
\sim \frac{\Lambda_{QCD}^2}{2m_Q},
\eeq
thus we expect $O_{\rm kin}$ to be
suppressed with respect to $D_t$ by a factor of $\Lambda_{QCD}/2m_Q$.
For the magnetic operator we have
\beq
\vec{B} = \vec{\nabla}\times \vec{A}.
\eeq
Since
\beq
A \sim p \sim O(\Lambda_{QCD})\;\mbox{ and } \nabla \sim
p \sim O(\Lambda_{QCD}),
\eeq
we estimate  the magnetic operator to also be of order
\beq
O_{\rm mag} \sim \frac{\Lambda_{QCD}^2}{2m_Q}.
\eeq

The heavy quark formulation used in this calculation is described in
more
detail in section~\ref{sec:simdet}, where we also talk briefly about
the
implementation we use for the light quark  and discuss the
choice of our simulation parameters. In section~\ref{sec:operators}
we show
the  operators from which
our meson and baryon states are constructed. Details of
our fitting procedure and the
results are discussed in section~\ref{sec:results}. We also
describe there how meson masses are extracted and
give results for the $B_d$ and the $B_s-B_d$ splitting. After that we
turn
to the $B^*-B$ (hyperfine) and $\Lambda_b-B$ splittings and compare
simulation results with experimental values and expectations from
HQET.
Finally, we present preliminary results for $P$ wave states. The
conclusions are
given in section~\ref{sec:conclusions}.
\section{\label{sec:simdet} Simulation details}
The gauge fields used here are an ensemble of 35 quenched
configurations at $\beta = 6.0$, generated by the UKQCD
collaboration. The configurations are separated by 200 sweeps, each
including 5 overrelaxation and 1 Cabibbo-Marinari heatbath
 sweep.  By time reversing them, we obtain
an additional set of
35 configurations which we treat in our analysis as statistically
independent. We fixed the configurations to Coulomb gauge.

In this simulation the non-relativistic Lagrangian, which describes
the $b$ quark, is expanded through order $1/m_Q^0$ at tree level:
\begin{equation}
{\cal L} = Q^{\dagger} \left( D_t +  H_0 + \delta H\right) Q,
\end{equation}
where
\begin{equation}
H_0 = -\frac{\vec{D}^2}{2m_{Q}^0},\; \delta H =
-\frac{c_B\,\vec{\sigma}\cdot\vec{ B}}{2m_Q^0}
\end{equation}
To get   rid of large renormalizations of the link operator, we use
tadpole
improvement~\cite{lepage93}:
\begin{equation}
U_{\mu} \rightarrow U_{\mu} /u_0, \;\; u_0^4 = \left\langle
{\scriptstyle
\frac{1}{3}} Tr
 U_{\rm Plaq.}\right\rangle ,\;\; u_0 \simeq 0.878,
\end{equation}
so we  can use the tree level coefficient $c_B=1$.
The heavy quark propagator is computed using the following
evolution equation~\cite{upsilon}:
\begin{eqnarray}
 G_1 &=&
  \left(1\!-\!\frac{aH_0}{2n}\right)^{n}
 U^\dagger_4
 \left(1\!-\!\frac{aH_0}{2n}\right)^{n} \, \delta_{\vec{x},0},\;t=1
\end{eqnarray}
and on the following timeslices
\begin{equation}
G_{t+1} = \left(1-\frac{aH_0}{2n}\right)^n U_4^{\dagger}
\left(1-\frac{aH_0}{2n}\right)^n (1-a\delta
H)G_t ,\; t>1,
\end{equation}
for  the heavy quark masses $am_Q^0$ = 1.71, 2.0, 2.5 and 4.0. The
bare heavy
quark mass corresponding to the $b$ quark should be the same as the
one tuned
in $\Upsilon$ spectroscopy calculations with NRQCD. If the
$b$ quark mass is set to $am_b^0 = 1.71$, the simulated
$\Upsilon$ mass agrees with the experimental value~\cite{upsilon},
using a
lattice spacing of 2.4(1) GeV, obtained from the $1P-1S$ and $2S-1S$
splittings
of the $\Upsilon$. However, due to the effect of quenching one
obtains
different lattice spacings from different physical quantities, or in
other
words, for different physical systems different scales are
appropriate. From
light hadron spectroscopy at the same value of $\beta$ one gets
lattice
spacings closer to $a^{-1} \sim 2$ GeV.
 Probably,  scales in  heavy-light systems are closer
 to the ones in light hadrons. In this paper we use 2.05(10) GeV. The
central
value is taken from the APE collaboration who performed a spectrum
calculation
at the same value of $\beta$ and the same type of light fermions
(clover
fermions with unity clover coefficient) and quote a value of $a^{-1}
=
2.05(6)$ GeV
from $m_{\rho}$~\cite{APE}.  We use a larger error for $a^{-1}$, to
encompass other results from light spectroscopy (e.g. $a^{-1} =
2.11(11)$ from
$f_{\pi}$ in~\cite{APE}, $a^{-1} = 2.0\err{3}{2}$ from $m_{\rho}$
in~\cite{kappac}.),
as well as from the string tension~\cite{bali} of $a^{-1} =1.94(8)$
GeV
 at $\beta$ = 6.0. Systematic errors in light hadron spectroscopy are
still
rather large and not well determined, so the uncertainty in $a^{-1}$
may be
underestimated. For $a^{-1} = 2.05$ GeV,
$am_Q^0 = 2.0$
corresponds to the bare $b$ quark mass used in the $\Upsilon$
simulations.
The stability coefficient $n$ has been set to 2 for all masses.

We also compute the
heavy quark propagator in the zeroth order (static) approximation of
the
$1/m^0_Q$ expansion, where the Lagrangian is just given by the
covariant
time derivative:
\beq
{\cal L}  = Q^{\dagger} D_t Q.
\eeq
and the heavy quark propagator follows the evolution equation:
\beq
G_{t+1} - U_4^{\dagger}G_t = \delta_{(\vec{x},t),0}.
\eeq

For the light quarks we use propagators generated by the UKQCD
collaboration
with a clover action~\cite{SW}:
\begin{eqnarray}
S_F &=& a^4
\sum_x\left[-\sum_{\mu}\frac{1}{a}
\kappa\left(\vphantom{A^1}\overline{q}(x)(1-\gamma_{\mu})U_{\mu}(x)
q(x+\mu)\right.\right. \nonumber \\
& &
+\left.\bar{q}(x+\mu)(1+\gamma_{\mu})U^{\dagger}_{\mu}(x)q(x)\right)
\nonumber \\
 & & \left.+\frac{1}{a}\bar{q}(x)
q(x)\vphantom{U^t}
-ica\kappa/2\sum_{\mu\nu}\bar{q}(x)
\sigma_{\mu\nu}F_{\mu\nu}q(x)\right],
\end{eqnarray}
with clover coefficient $c=1$, i.e. {\em without} tadpole
improvement.
The quark fields are rotated at the source:
\beq
q(x) \rightarrow \left(1-\frac{a}{2}\gamma\cdot D\right)q(x).
\eeq
It has been shown~\cite{Heatlie91} that matrix elements calculated
with these
quark fields are free of $O(a)$ errors at tree level.

We use light fermions at values of $\kappa$ of $0.1440$ and $0.1432$,
which
bracket the strange quark mass. On the same set of configurations and
light propagators the critical $\kappa$ values has been
determined to be $0.14556(6)$. The strange quark, whose mass is
determined
from $m_K^2/m_{\rho}^2$, corresponds to
$\kappa_s = 0.1437\errr{4}{5}$~\cite{kappac}.
\section{\label{sec:operators} Hadron operators}
Our meson states are built up from  various interpolating operators:
\beq
\sum_{\vec{x}_1,\vec{x}_2}{Q}^{\dagger}(\vec{x}_1)\Gamma^{\dagger}
(\vec{x}_1 -\vec{x}_2) q(\vec{x}_2).
\eeq
 $\Gamma(\vec{x}_1-\vec{x}_2)$ factorizes into a smearing
function $\phi(r=|\vec{x}_1 - \vec{x}_2|)$ and an operator $\Omega$
which consists of a $4\times 2$ matrix in spinor space and, for $P$
states, a derivative acting on $\phi$. We choose $\phi$ to be either
a
delta function or, to project onto the
ground or an excited state, a hydrogen-like wave function. The
smearing is
applied on
the heavy quark. In general the
smearing function $\phi$ is
different at the source $(sc)$  and at the sink $(sk)$, and we obtain
meson propagators of the following form:
\beq
C(\vec{p}=0,t)= \sum_{\vec{y}_1,\vec{y}_2} Tr
\left[\gamma_5(M^{-1})^{\dagger}(\vec{y}_2)\gamma_5
\Gamma^{(sk)\dagger}(\vec{y}_1-\vec{y}_2)
\tilde{G}_t(\vec{y}_1)\right].
\eeq
Here, $M^{-1}$ is the light quark propagator and
\beq
\tilde{G}_t(\vec{y}) = \sum_{\vec{x}}G_t(\vec{y} - \vec{x})
\Gamma^{(sc)}
(\vec{x})
\eeq
is the heavy quark propagator smeared at the source.

\begin{table}
\begin{center}
\begin{tabular}{|ccc|}
\hline
Meson &Lattice  &  \\
$^{2S+1}L_J$ ($J^{P}$)&  Rep.  & $\Omega$  \\
\cline{1-3}
&& \\
 ${^1S}_0\;(0^{-})$ &  $A^{-}_1$  &$\left(\begin{array}{c}0\\1\!\!1
\end{array}\right)$ \\
&& \\
 ${^3S}_1\;(1^{-})$ &  $T^{-}_{1(i)}$  & $\left(\begin{array}{c}0\\
\sigma_i\end{array}\right) $    \\
&& \\
 ${^1P}_1\;(1^{+})$ &  $T^{+}_{1(i)}$  & $\left(\begin{array}{c}0\\
\Delta_i\end{array}\right) $  \\
&& \\
 ${^3P}_0\;(0^{+})$ &  $A^{+}_1$  & $\left(\begin{array}{c}0\\
\sum_j\Delta_j  \sigma_j \end{array}\right) $  \\
&& \\
 ${^3P}_1\;(1^{+})$ &  $T^{+}_{1(ij)}$  & $ \left(\begin{array}{c}0\\
\Delta_i \sigma_j - \Delta_j
\sigma_i \end{array}\right)$  \\
&& \\
 ${^3P}_2\;(2^{+})$ &  $E^{+}_{(ij)}$  & $ \left(\begin{array}{c}0\\
\Delta_i \sigma_i - \Delta_j
\sigma_j\end{array}\right)$  \\
&& \\
                     &  $T^{+}_{2(ij)}$  & $
\left(\begin{array}{c}0\\
\Delta_i \sigma_j + \Delta_j
\sigma_i\end{array}\right)$  \\
&&$\qquad$ ($i \neq j$) \\
&& \\
\cline{1-3}
\end{tabular}
\end{center}
\caption{Meson Operators. $\Delta_i$ denotes the symmetric lattice
derivative. $1\!\!1$ stands for the $2\times 2$ unit matrix.}
\label{table:operators}
\end{table}
The
continuum quantum numbers of the  $S$ and $P$ states  we implemented
and the corresponding lattice operators $\Omega$  can be
found in
table~\ref{table:operators}. For the $\gamma$ matrices we chose the
Dirac basis
using the representation
\beq
\gamma_0 =\left(\begin{array}{lr}1\!\!1& 0 \\ 0 &-1\!\!1
\end{array}\right) ,\;\;\vec{\gamma} =
\left(\begin{array}{lr}0&-i\vec{\sigma}
\\i\vec{\sigma}& 0
\end{array}\right).
\eeq
Note that in heavy-light systems $C$ is
not a good  quantum number, and we expect mixing between the $^3P_1$
and
the $^1P_1$ states.

For the $\Lambda_b$ baryon we used the following interpolating
operator:
\beq
O_{\alpha} = \epsilon_{abc} \sum_{\vec{x}_1} Q_{\alpha}^a(\vec{x}_1)
\sum_{\vec{x}_2}(q^b)^T(\vec{x}_2)C\gamma_5\gamma_0
q^c(\vec{x}_2)\phi(
|\vec{x}_1-\vec{x}_2|),
\eeq
where $C = \gamma_0 \gamma_2$ is the charge conjugation matrix.
The baryon correlators are smeared at the source and local at the
sink.
\section{\label{sec:results}Results}
In the following we will use the notation $C_{rs}$ for the
correlation
functions, where the index $r$ denotes a delta function $(L)$, ground
state
smearing function $(1)$ or excited state smearing function $(2)$ at
the source.
The index $s$ denotes the  smearing function at the sink.

{}From the original data we generated 100 bootstrap ensembles, each
containing
the original number of samples, and fitted to each of them
separately. The
fit results and errors are obtained from averaging over the number of
bootstrap samples. This procedure also enables us to take
correlations
between data
with different $\kappa$ and $m_Q^0$ values from the same
configuration into
account. Given the relatively low statistics, we are sometimes forced
to
discard eigenvalues in our SVD inversion of the covariance matrix.
This is
done using a cutoff $\lambda$ on the ratio between the smallest and
the largest
eigenvalue.
\subsection{Ground state energy}
To extract the bare ground state energy, $E_{\rm sim}$,
and amplitudes we fitted the correlators
$C_{1L}$ and $C_{11}$ simultaneously to a single exponential. The
signal is good up to
large times and we fit out until $t_{max}=25$. The plateau of
the correlation functions at $\kappa$ = 0.1432 is  reached at
$t_{min} = 5$,
but at $\kappa$ = 0.1440  there is
a slight decrease in the ground state energy and especially the
amplitudes
until $t_{min}$
is moved out to 9. Moving $t_{min}$ further out does not change the
results
any more.  An example for the dependence of the ground state energy
on the fit interval at the different
$\kappa$ values is shown in table~\ref{table:1exp1}, together with
the
quality of fit parameter $Q$.

With our limited statistics, our bootstrap procedure generates
certain
 ensembles on which
 multi-exponential fits fail, but with the original ensemble of
correlation
functions we
can do a simultaneous fit of $C_{1L}$ and $C_{2L}$ to 2 exponentials.
This gives a value for the ground state energy which tends to be a
little
higher, but in general still compatible within one
standard deviation with the
results fitted with just one exponential. Results of double
exponential fits
for $am_Q^0=4.0$ are shown in table~\ref{table:2exp}.
The single exponential fits give higher $Q$ values and better looking
effective mass plots, so
we  use them  to extract the meson ground state
energies. We can also make a rough estimate of the energy of the
first
excited state. Its error bars are too large to
extract a dependence on the heavy or the light quark masses. From the
results
in table~\ref{table:2exp} and fit results at other heavy quark masses
we
take the value of
$E_{\rm sim}(2S) = 0.8(1)$ as a reasonable estimate. For a more
reliable
calculation of the excited state energy one would have to do three
exponential fits to $C_{1L}$ and $C_{2L}$, but, given our low
statistics,
errors on these fits are very large.
\begin{table}
\begin{center}
\begin{tabular}{|r|c|c|c|c|}
\hline
                   \multicolumn{1}{|c|}{}
                 & \multicolumn{2}{c|}{$\kappa=0.1432$}
                 & \multicolumn{2}{c|}{$\kappa=0.1440$} \\
\hline
 \multicolumn{1}{|r|}{$t_{min}/t_{max}$}
 & \multicolumn{1}{c}{$aE_{\rm sim}(1S)$}
 & \multicolumn{1}{c}{Q}
 & \multicolumn{1}{|c}{$aE_{\rm sim}(1S)$}
 &  \multicolumn{1}{c|}{Q} \\
\hline
 $5/25$  & 0.516(6) &0.3 & $0.497(6)$& $0.2$\\
 $6/25$  & 0.515(7) &0.4 & $0.495(7)$& $0.2$ \\
 $7/25$  & 0.514(6) &0.4 & $0.493(8)$& $0.2$  \\
 $8/25$  & 0.514(7) &0.4 & $0.491(7)$& $0.3$  \\
  $9/25$ & 0.513(7) &0.4 & $0.491(8)$& $0.3$  \\
 $10/25$ & 0.511(5) &0.4 & $0.489(7)$ &$0.3$   \\
\hline
\end{tabular}
\end{center}
\caption{Results of single exponential fits to the ground state
correlation
functions $C_{11}$ and  $C_{1L}$ at $am_Q^0$ = $4.0$. For $\kappa =
0.1432$,
$\lambda = 0.03$, for $\kappa = 0.1440$, $\lambda=0.02$.}
\label{table:1exp1}
\end{table}

\begin{figure}[pthb]
\vspace{-2cm}
\centerline{\epsfxsize=10cm
\epsfbox{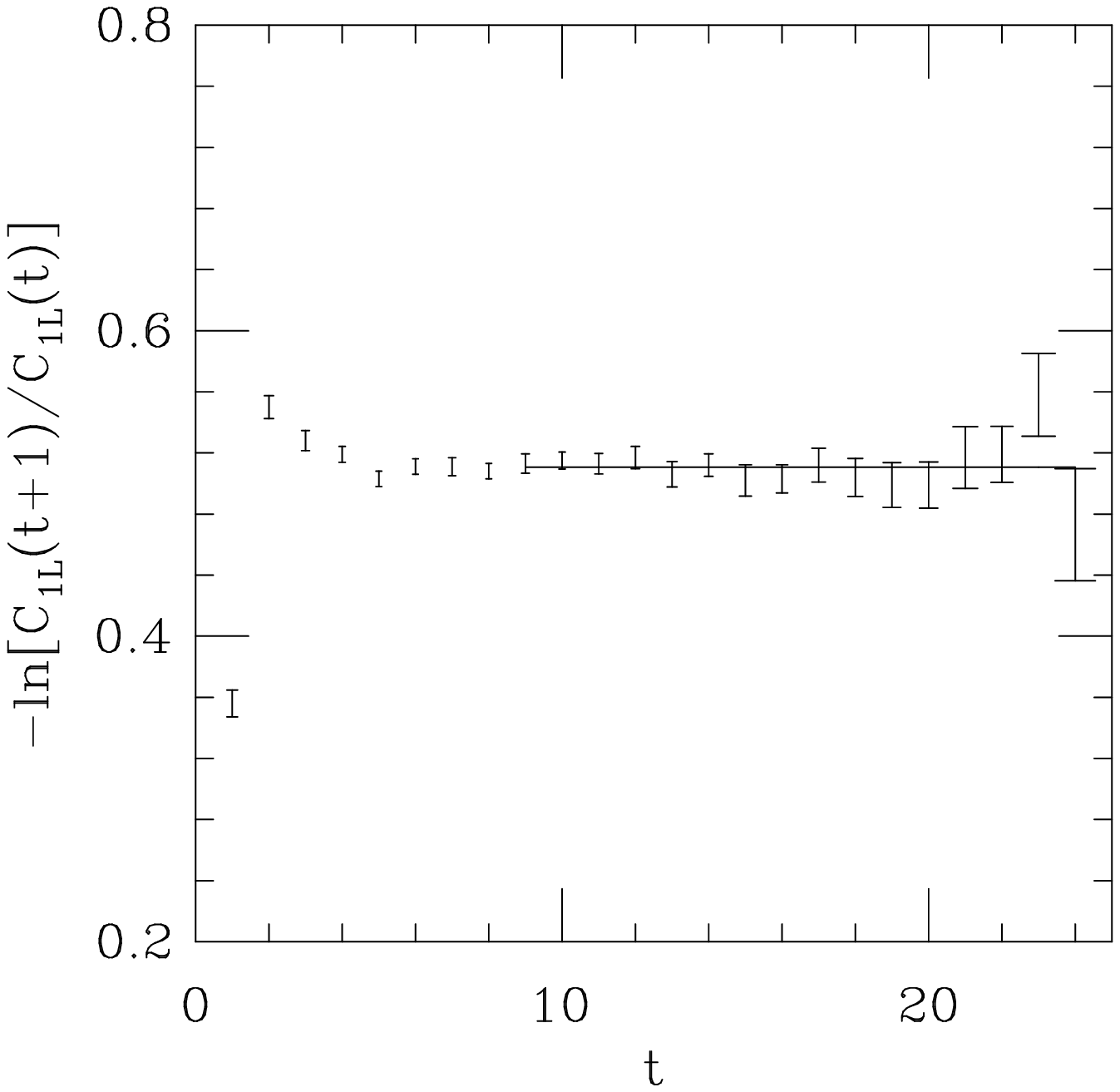}
\epsfxsize=10cm
\epsfbox{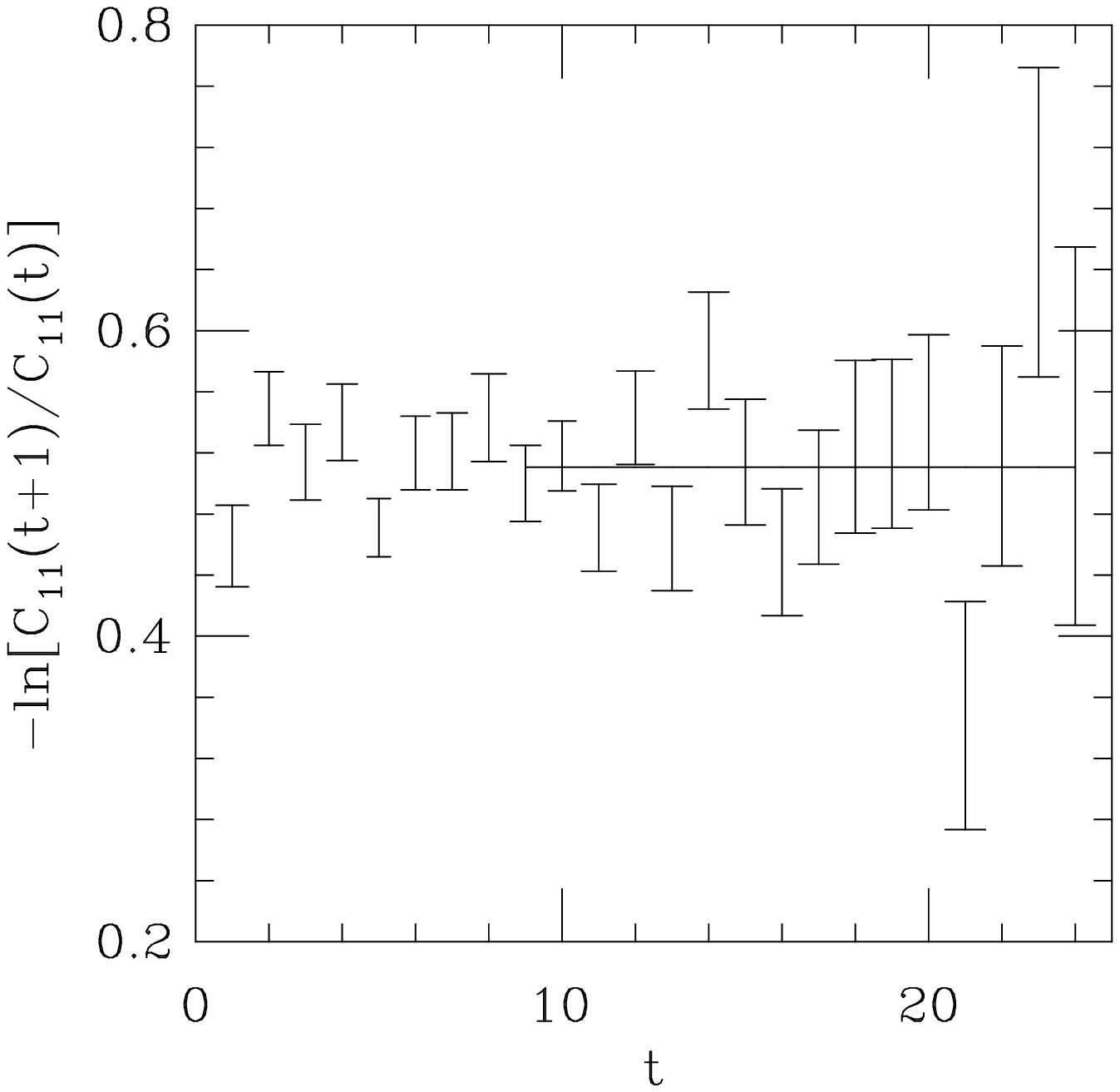}}
  \caption{Effective masses of $C_{1L}$ (left) and $C_{11}$ (right)
correlation functions at $am_Q^0 = 2.5$ and $\kappa = 0.1432$. }
 \label{fig:effmass}
\end{figure}

We extract meson energies from fits in the range
$t_{min}/t_{max} = 9/25$ for both $\kappa$ values. The systematic
error due to
the variation of the fit interval and the difference between the
single and
double exponential fits is smaller than the fit error at this fit
range.
 Effective mass plots for $C_{1L}$
and $C_{11}$ are shown in figure~\ref{fig:effmass}. Results for the
pseudoscalar, vector and spin-averaged ground state energy
$\overline{E} = (3E_{\rm sim}(^3S_1)+E_{\rm sim}(^1S_0))/4$ for all
heavy
and light quark mass values are shown in
table~\ref{table:fit_ground_state},
the chiral extrapolation and interpolation to the strange
light quark mass in table~\ref{table:extrap_ground_state}.
\begin{table}
\begin{center}
\begin{tabular}{|r|l|l|c|l|l|c|}
\hline
                   \multicolumn{1}{|c|}{}
                 & \multicolumn{3}{c|}{$\kappa=0.1432$}
                 & \multicolumn{3}{c|}{$\kappa=0.1440$} \\
\hline
 \multicolumn{1}{|r|}{$t_{min}/t_{max}$}
 & \multicolumn{1}{c}{$aE_{\rm sim}(1S)$}
 & \multicolumn{1}{c}{$aE_{\rm sim}(2S)$}
 & \multicolumn{1}{c}{$Q$}
 & \multicolumn{1}{|c}{$aE_{\rm sim}(1S)$}
 & \multicolumn{1}{c}{$aE_{\rm sim}(2S)$}
 & \multicolumn{1}{c|}{$Q$} \\
\hline
 $4/24$ & $0.513(4)$  & $0.81(4)$ & $0.07$ & $0.495(6)$ & $0.79(5 )$&
$0.09$ \\
 $5/25$ & $0.518(5)$  & $0.77(6 )$& $0.17$ & $0.497(7)$ & $0.71(6 )$&
$0.19$\\
 $6/25$ & $0.516(5)$  & $0.79(8 )$& $0.15$ & $0.495(9)$ & $0.71(9 )$&
$0.15$ \\
 $7/25$ & $0.519(3)$  & $0.94(34)$& $0.09$ & $0.502(5)$ & $0.92(25)$&
$0.08$ \\
 $8/25$ & $0.520(7)$  & $0.91(18)$& $0.07$ & $0.498(3)$ & $0.92(43)$&
$0.11$ \\
 $9/25$ & $0.517(7 )$ & $0.90(18)$& $0.06$ & $0.501(5)$ & $0.88(25)$&
$0.09$ \\
 $10/25$& $0.517(6)$  & $0.88(16)$& $0.03$ & $0.501(2)$ & $0.86(8 )$&
$0.07$ \\
\hline
\end{tabular}
\end{center}
\caption{Results of double  exponential fits to the correlation
functions
$C_{1L}$ and $C_{2L}$ for $am_Q^0$ = 4.0,
$\lambda = 10^{-5}$.}
\label{table:2exp}
\end{table}
\begin{table}
\begin{center}
\begin{tabular}{|c|c|c|c|c|c|}
\hline
                   \multicolumn{2}{|c|}{}
                 & \multicolumn{2}{c|}{$\kappa=0.1432$}
                 & \multicolumn{2}{c|}{$\kappa=0.1440$} \\
\hline
$^{3S+1}L_J$& $ am_Q^0$&$\;\;a E_{\rm sim} $&Q&$aE_{\rm sim}$& Q\\
\cline{1-6}
$^1S_0$ & 1.71   & 0.501(9) &0.3  & 0.480(7) &0.4  \\
        & 2.0    & 0.506(6) & 0.4 & 0.483(7) &0.4  \\
        & 2.5    & 0.511(5) & 0.4 & 0.487(7) &0.4  \\
        & 4.0    & 0.513(7) & 0.3 & 0.491(8) &0.3  \\
       &$\infty$ & 0.524(6) & 0.2 & 0.508(7) &0.2  \\
\hline
$^3S_1$ & 1.71  & 0.521(6) &0.3   & 0.502(7) &0.4  \\
        & 2.0   & 0.523(5) & 0.3  & 0.503(6) &0.3  \\
        & 2.5   & 0.524(6) & 0.3  & 0.504(6) &0.3  \\
        & 4.0   & 0.527(6) & 0.3  & 0.507(8) &0.2  \\
\hline
spin avg. & 1.71  & 0.516(6) &0.3   & 0.497(7) & \\
          & 2.0   & 0.518(5) & 0.3  & 0.498(7) &  \\
          & 2.5   & 0.521(6) & 0.3  & 0.499(7) &  \\
          & 4.0   & 0.524(6) & 0.3  & 0.503(7) &  \\
\hline
\end{tabular}
\end{center}
\caption{Fit results for the bare ground state energy for the
pseudoscalar
and the vector meson and the spin-averaged energy for both $\kappa$
values.}
\label{table:fit_ground_state}
\end{table}
\begin{table}
\begin{center}
\begin{tabular}{|c|c|l|c|}
\hline
                   \multicolumn{2}{|c|}{}
                 & \multicolumn{1}{l|}{$m_q = 0$}
                 & \multicolumn{1}{l|}{$m_q = m_{\rm strange}$} \\
\hline
$^{3S+1}L_J$&$ am_Q^0$&$\;\;a E_{\rm sim} $ &$aE_{\rm sim}$ \\
\cline{1-4}
$^1S_0$         & 1.71 & 0.437(19)&    0.489(6) \\
                & 2.0  & 0.437(11)&    0.491(6)    \\
                &2.5  &  0.439(12)&     0.496(7) \\
                &4.0  &  0.446(15)&     0.499(6) \\
                &$\infty$ & 0.461(15)& 0.511(6) \\
\cline{1-4}
$^3S_1$         & 1.71   &$0.464(15)$ & $0.509(6)$  \\
                & 2.0    &$0.463(12)$ & $0.510(6)$   \\
                &2.5     &$0.464(9)  $    & $0.512(6)$\\
                &4.0     &$0.478(12) $& $0.515(6)$  \\
\hline
spin avg.       & 1.71   &$0.457(14)$ & $0.504(6)$  \\
                & 2.0    &$0.456(12)$ & $0.505(6)$   \\
                &2.5     &$0.457(9)  $    & $0.508(6)$\\
                &4.0     &$0.462(13) $& $0.511(6)$  \\
\hline
\end{tabular}
\end{center}
\caption{Bare ground state energy, extrapolated to a vanishing (left)
and the
strange (right) light quark mass. }
\label{table:extrap_ground_state}
\end{table}

\begin{figure}[hbtp]
\vspace{-4cm}
\centerline{\ewxy{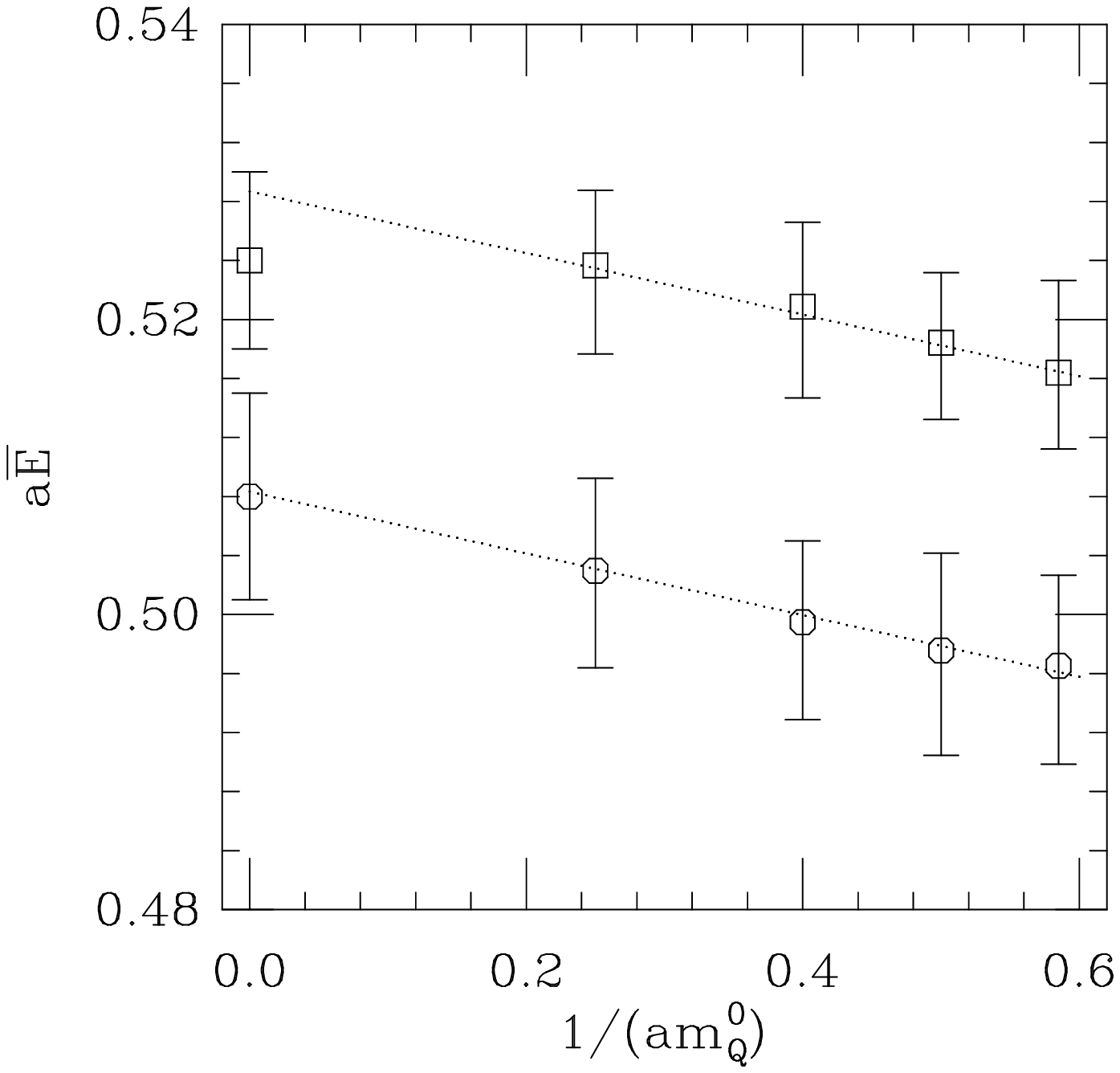}{13cm}}
  \caption{Bare spin-averaged ground state energy as a function of
the
inverse bare heavy quark mass $am_Q^0$. Squares denote $\kappa =
0.1432$
and circles $\kappa=0.1440$. The lines represent correlated fits of
the
NRQCD results to a linear function.}
 \label{fig:E_sim}
\end{figure}

The dependence of the bare spin-averaged ground state energy on the
bare
heavy quark mass is illustrated in figure~\ref{fig:E_sim}. This
quantity is
spin-independent and thus useful to estimate the correction due  to
the
kinetic operator $O_{\rm kin}$. Correlated fits of the simulation
results
to linear functions give a slope of $-0.02(2)$
in lattice units for $\kappa = 0.1440$ and for $\kappa = 0.1432$,
$-0.02(1)$. In both cases, the extrapolation to the static limit is
compatible with the static simulation results. We note that in the
tadpole
improved case,
$O_{\rm kin}$ is not a positive definite operator. A more physical
quantity
than $E_{\rm sim}$ is the binding energy,
$E_{\rm bind}=E_{\rm sim}-E_0$, where $E_0$ is the shift in the zero
of energy for NRQCD or the static theory, respectively
(perturbative values from~\cite{colinII}
 listed in table~\ref{table:M_P}). This quantity
rises slightly  as a function of $1/m_Q^0$, but there are significant
perturbative uncertainties. For a more detailed discussion
see~\cite{spec_dynamical}.
\subsection{\label{mes_mass}Meson masses}
  We calculate the meson mass using
 the relation
\beq
M_B = \Delta + E_{\rm sim}.
\eeq
The energy shift $\Delta$ can be determined in perturbation theory.
It consists
of the renormalized heavy quark mass and a shift in the zero of
energy:
\beq
\Delta = Z_m m_Q^0 - E_0.
\eeq
The quantities $Z_m$ and $E_0$ are calculated at $O(\alpha)$:
\beq
aE_0 = A\alpha_V(q^{*}_A),\;Z_m = 1 + B \alpha_V(q^{*}_B),
\eeq
where we take for $\alpha_V$ the two loop expression for the running
coupling
constant in the Lepage-Mackenzie
scheme. The one loop terms $A$ and $B$  and the appropriate momentum
scales
$q^{*}_A$ and $q^{*}_B$ were calculated by
C.~Morningstar~\cite{colinII}. The shifts used in our calculation are
shown in
table~\ref{table:M_P}, together with
results for the meson masses in lattice and physical units.  $E_0$
values are
perturbative, shifts are perturbative except in the last row.
At $m_Q^0=4.0$, a  non-perturbative
energy shift calculated from the $\Upsilon$ has been used, to avoid
defects
in the scale $q^*_B$ in the heavy quark mass renormalization,
occuring at heavy quark masses around $am_Q^0 = 5$.
\begin{table}
\begin{center}
\begin{tabular}{|l|c|c|c|c|c|c|c|}
\hline
                 \multicolumn{1}{|c|}{}
                & \multicolumn{1}{c|}{}
                & \multicolumn{1}{c|}{}
                & \multicolumn{1}{c}{}
                 & \multicolumn{3}{|c|}{$aM_P$}
                 & \multicolumn{1}{c|}{} \\
\cline{5-7}
                   \multicolumn{1}{|c}{$am_Q^0$}
                 & \multicolumn{1}{|c}{$n$}
                 & \multicolumn{1}{|c}{$aE_0$}
                 & \multicolumn{1}{|c|}{$a\Delta$}
                 & \multicolumn{1}{c|}{$\kappa=0.1432$}
                 & \multicolumn{1}{c|}{$\kappa=0.1440$}
                 & \multicolumn{1}{c|}{$m_q=0$}
                 & \multicolumn{1}{c|}{$M_{P_d}$[GeV]} \\
\hline
 $1.71$ & $2$ & $0.32(6)$ & $1.73(13)$  &$2.23$  & $2.21$& $2.17$&
$4.5(5)$  \\
 $2.0$ &  $2$ & $0.30(6)$ & $2.02(12)$  &$2.53$  & $2.50$& $2.46$ &
$5.1(5)$\\
 $2.5$&   $2$ & $0.29(6)$ & $2.48(9)$   &$2.99$  & $2.97$& $2.92$ &
$6.0(5)$\\
 $4.0$ &  $2$ &           & $4.07(5)$   &$4.58$  & $4.56$& $4.52$ &
$9.3(6)$ \\
\hline
\end{tabular}
\end{center}
\caption{Energy shifts used in our calculation
and pseudoscalar meson masses calculated with these
shifts.  The last
column contains meson masses for $m_q\rightarrow 0$
in physical units, using a lattice spacing of $a^{-1} =2.05(10)$
GeV. Errors on $aM_P$ are dominated by the
perturbative error in $\Delta$, errors on the physical masses by
perturbation theory and the uncertainty in the lattice spacing.}
\label{table:M_P}
\end{table}
\begin{table}
\begin{center}
\begin{tabular}{|l|lcl|lcl|}
\hline
                   \multicolumn{1}{|l|}{}
                 & \multicolumn{3}{c|}{$\kappa=0.1432$}
                 & \multicolumn{3}{c|}{$\kappa=0.1440$} \\
\hline
                 \multicolumn{1}{|c}{$am_Q^0$}
 & \multicolumn{1}{|l}{$a\Delta E$}
 & \multicolumn{1}{c}{Q}
 & \multicolumn{1}{l|}{$aM_P$}
 & \multicolumn{1}{l}{$a\Delta E$}
 & \multicolumn{1}{c}{Q}
 & \multicolumn{1}{l|}{$aM_P$} \\
\hline
$1.71$ & $0.030(7) $  & $0.5$ & $2.6(6)  $  & $0.034(5)$   & $0.5$ &
$2.3(3)$     \\
$2.0$  & $0.030(12)$  & $0.5$ & $2.6(1.0)$& $0.030(5)$   & $0.5$ &
$2.6(4)$   \\
$2.5$  & $0.024(8)$   & $0.5$ & $3.2(1.1)$  & $0.026(5)$   & $0.5$ &
$3.0(6)$    \\
$4.0$  & $0.018(8)$   & $0.4$ & $4.3(1.9)$  & $0.018(5)$   & $0.5$ &
$4.3(1.2)$   \\
\hline
\end{tabular}
\end{center}
\caption{Splitting between correlation functions with
$|\vec{p}|=2\pi/L_s$ and
$|\vec{p}|=0$ and meson masses derived from it.}
\label{table:kinetic_mass}
\end{table}
It is possible to calculate the meson mass non-perturbatively from
the meson dispersion relation~\cite{MBpaper}. The non-relativistic
dispersion relation reads in lowest order:
\beq
E(\vec{p}) = E(\vec{p} = \vec{0}) + \frac{\vec{p}^2}{2M_B}.
\eeq
Note that $E(\vec{p} = 0)$ is, by definition, $E_{\rm sim}$.
We calculate the `kinetic' meson mass $M_B$ from the
energy splitting between mesons of momentum $|\vec{p}| = 2\pi/L_s$
and
$|\vec{p}| = 0$, where $L_s$ is the spatial lattice extent.

In our analysis of the energy splittings between the finite and zero
momentum correlation functions we make use of the fact that
correlation
functions of different states on the same configuration are highly
correlated.
We generate 100 bootstrap measurements of the jackknifed ratio $R$ of
the
correlation functions  of the two states and fit these to a single
exponential:
\beq
R \sim A e^{-\Delta E  t},
\eeq
where $\Delta E$ is the energy splitting.

As shown in
table~\ref{table:kinetic_mass}, the masses obtained with this method
have much larger errors, but are compatible with the masses
calculated
using perturbation theory.
\subsection{Mass splittings}
\subsubsection{$B_s-B_d$ splitting}
\begin{table}[tphb]
\begin{center}
\begin{tabular}{|l|l|}
\hline
 $am_Q^0$&$\;\;a \Delta E \;\;$\\
\cline{1-2}
$1.71$ & 0.047(11)  \\
$2.0$  & 0.049(8)  \\
$2.5$  &  0.050(5)  \\
$4.0$  &  0.049(7) \\
$\infty$ & 0.051(13) \\
\hline
\end{tabular}
\end{center}
\caption{Spin-averaged $B_s-B_d$ splitting.}
\label{table:s-d}
\end{table}
\begin{figure}[bhtp]
\vspace{-3cm}
\centerline{\ewxy{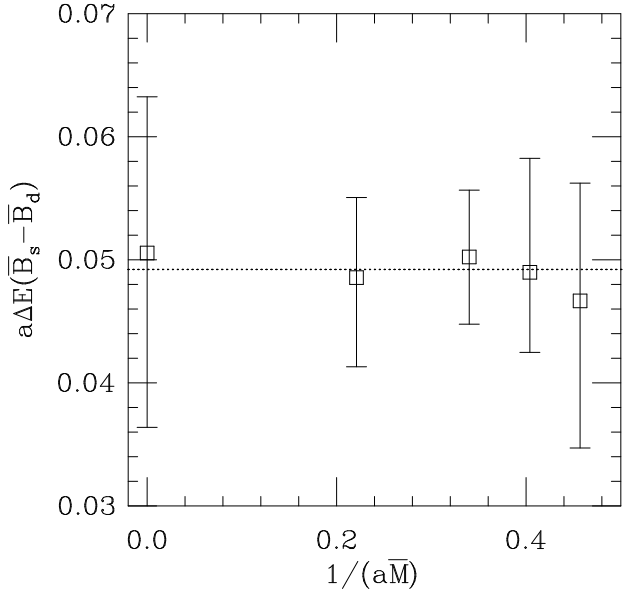}{13cm}}
  \caption{Spin-averaged energy splitting between $B_s$ and $B_d$ in
lattice
units plotted against the inverse spin-averaged meson mass.
The dotted line denotes a correlated fit of the simulation results to
a
constant.}
 \label{fig:s-d}
\end{figure}
Results for the spin-averaged $B_s-B_d$ splitting are listed in
table~\ref{table:s-d}. As expected from HQET, this splitting is
fairly
independent of the heavy quark mass. Figure~\ref{fig:s-d} shows the
splitting as a function of the inverse spin-averaged meson mass  with
a correlated fit to a constant. The fitted value for this constant is
 $0.049(5)$, with a $\chi^2$ per degree of freedom of $0.03$.
Converted into physical units this  is $100(14)$ MeV, which is
is in good agreement with the experimental value of $97(6)$ MeV. This
also
means that $\kappa_s$ as determined from the $K$ mass is appropriate
to the
$B_s$ system.
\subsubsection{Hyperfine splitting}
\begin{figure}[bhtp]
\vspace{-2cm}
\centerline{\ewxy{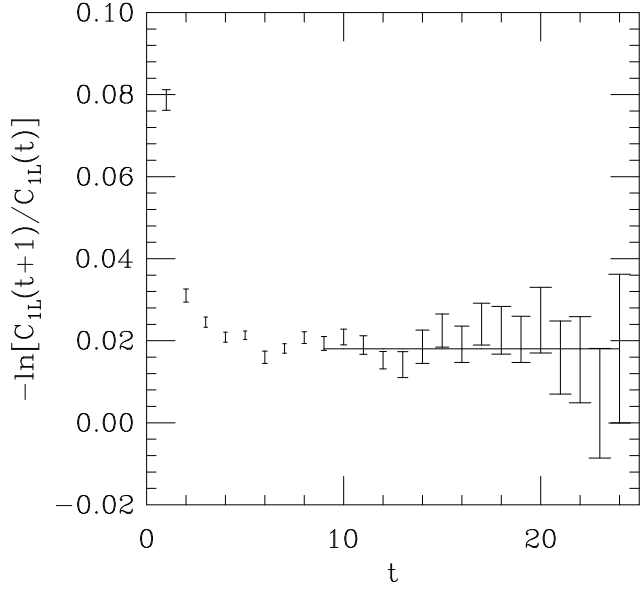}{13cm}}
  \caption{Effective mass of the ratio of the correlation functions
of the
$B^*$ and the $B$ at $am_Q^0$ = 2.0, $\kappa$ = 0.1432.
 The solid line denotes the fit. }
 \label{figure:effmass_B*_B}
\end{figure}

For the hyperfine splitting we use the ratio fit method as described
in section
\ref{mes_mass} for the splitting between finite and zero momentum
correlation functions. The fit interval is chosen to be
$t_{min}/t_{max}$ = $9/25$.
An effective mass plot is shown in figure~\ref{figure:effmass_B*_B}.
The fit
results are given in table~\ref{table:bstarb_fit}. There is no
visible
dependence on the light quark mass. This also holds for the
experimental
results, which are $46.0(6)$ MeV for the  $B^*-B$ splitting and
$47.0(2.6)$ MeV  for the $B_s ^*-B_s$ splitting.

\begin{table}
\begin{center}
\begin{tabular}{|c|c|c|c|c|}
\hline
                   \multicolumn{1}{|c|}{}
                 & \multicolumn{2}{c|}{$\kappa=0.1432$}
                 & \multicolumn{2}{c|}{$\kappa=0.1440$} \\
\hline
$am_Q^0$&$\;\;a\Delta E \;\;$&Q&$\;\;a\Delta E \;\;$&Q\\
\cline{1-5}
1.71    & 0.0201(17) &0.43  & 0.0200(24) &0.42  \\
2.0    & 0.0180(14) & 0.51  & 0.0180(20) &0.47  \\
2.5   & 0.0151(12) & 0.43   & 0.0150(17) &0.37 \\
4.0    & 0.0103(10) & 0.28  & 0.0102(14) &0.22  \\
\cline{1-5}
\end{tabular}
\end{center}
\caption{Fit results for the $B^* - B$ splitting in lattice units,
$\lambda = 5\times 10^{-2}$.}
\label{table:bstarb_fit}
\end{table}
\begin{figure}[hbtp]
\vspace{-4cm}
\centerline{\ewxy{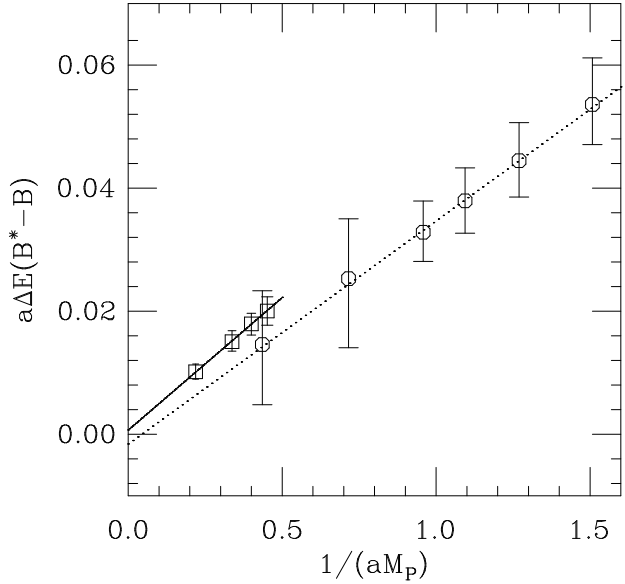}{13cm}}
  \caption{$B^*-B$ splitting at $\kappa = 0.1440$
plotted against the
inverse pseudoscalar meson mass. Squares denote NRQCD results and
circles,
UKQCD results using the clover formulation for the heavy quark,
plotted
against the
inverse dynamical meson mass~\protect\cite{sara}.
The dotted lines show  correlated fits of
the data to linear  functions. Errors in the meson mass are not shown
for
clarity. }
 \label{fig:Bstar_corr3}
\end{figure}

We find our simulation results for the splitting to be
approximately proportional to the heavy meson mass:
\begin{equation}
\Delta E \propto 1/M_P.
\end{equation}
This is  illustrated in figure~\ref{fig:Bstar_corr3}. Using a
correlated fit
 to the
results at $\kappa = 0.1440$ it is found that the splitting
extrapolates to
$a\Delta E = 0.001(2)$ as $M_P \rightarrow \infty$. This is
consistent with zero, as it should be. The fit
result for the slope is  $0.043(6)$. From a fit to the simulation
results at  $\kappa = 0.1432$ one obtains very similar values for the
static extrapolation and the slope.
Converting the  value corresponding to the physical $B$ mass
 into physical units, we obtain 36 MeV. This is relatively close  to
the
experimental value, but slightly low. This might be an
effect of quenching, which is expected to decrease the wave function
at the
origin and thus the hyperfine splitting. The error is 5 MeV if
uncertainties
in the lattice spacing are not taken into account; if they are
included one
obtains an error of 9 MeV.  Note that errors in $a^{-1}$ effectively
appear
squared in
the hyperfine splitting, as they affect both the value of the
splitting and
the meson mass at which the splitting is taken. Previous
studies~\cite{ukqcd_old} by UKQCD using a relativistic quark action
for the
heavy quark have given much lower results, which are incompatible
with
experiment, when plotted against the `static' meson mass. As we show
in
figure~\ref{fig:Bstar_corr3}, their results, using the same ensemble
of gauge
field configurations and light propagators and clover
fermions without tadpole improvement for the heavy quark, are
compatible
with ours when plotted against the kinetic mass~\cite{sara}.
However there is a systematic uncertainty in the determination of the
kinetic
mass for clover fermions of about 20 \% around the $B$, as well as
statistical errors that are  larger than 30 \% in the region of the
$B$.
\subsubsection{$\Lambda_b - \overline{B}$ and $\Lambda_b - B$
splittings}
\begin{table}
\begin{center}
\begin{tabular}{|c|c|c|c|c|}
\hline
                   \multicolumn{1}{|c|}{}
                 & \multicolumn{2}{c|}{$\kappa=0.1432$}
                 & \multicolumn{2}{c|}{$\kappa=0.1440$} \\
\hline
$am_Q^0$&$\;\;aE_{\rm sim} \;\;$&Q&$\;\;a E_{\rm sim} \;\;$&Q\\
\cline{1-5}
1.71  & 0.82(1) & 0.25  & 0.76(2) &0.41  \\
2.0   & 0.82(1) & 0.31  & 0.76(2) &0.43  \\
2.5   & 0.82(2) & 0.31  & 0.77(2) &0.47 \\
4.0   & 0.82(2) & 0.34  & 0.77(2) &0.40  \\
\cline{1-5}
\end{tabular}
\end{center}
\caption{Fit results for the $\Lambda_b$ ground state energy in
lattice units.}
\label{table:fit_lambda_b_B}
\begin{center}
\renewcommand{\arraystretch}{1.1}
\begin{tabular}{|l|c|c|}
\hline
$am_Q^0$&$\;\;a\Delta E(\Lambda_b-B) \;\;$&$a\Delta
E(\Lambda_b-\overline{B})$ \\
\cline{1-3}
1.71 & 0.21(7)& 0.19(6)  \\
2.0  & 0.22(4)& 0.20(5) \\
2.5  &  0.22(4)& 0.20(4) \\
4.0  &  0.21(4)& 0.20(5)\\
\cline{1-3}
\end{tabular}
\end{center}
\caption{Chirally extrapolated splitting between the $\Lambda_b$ and
the $B$
and between the $\Lambda_b$ and the
spin average between the $B$ and $B^*$, in lattice units.}
\label{table:res_lambda_b_B}
\end{table}
\renewcommand{\arraystretch}{1.0}
\begin{figure}[tbhp]
\vspace{-2cm}
\centerline{\ewxy{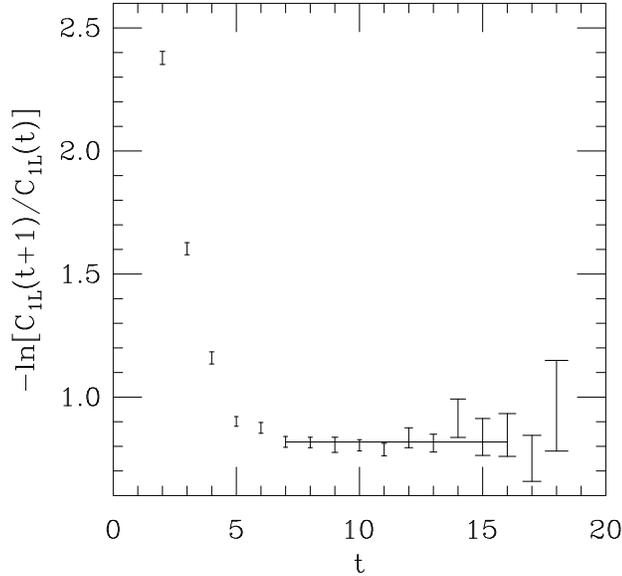}{13cm}}
  \caption{Effective mass of the $\Lambda_b$ correlation function
 at $am_Q^0$ = 2.0, $\kappa$ = 0.1432. The solid line
denotes the fit. }
 \label{figure:effmass_Lb_B}
\end{figure}
\begin{figure}[bhtp]
\vspace{-4cm}
\centerline{\ewxy{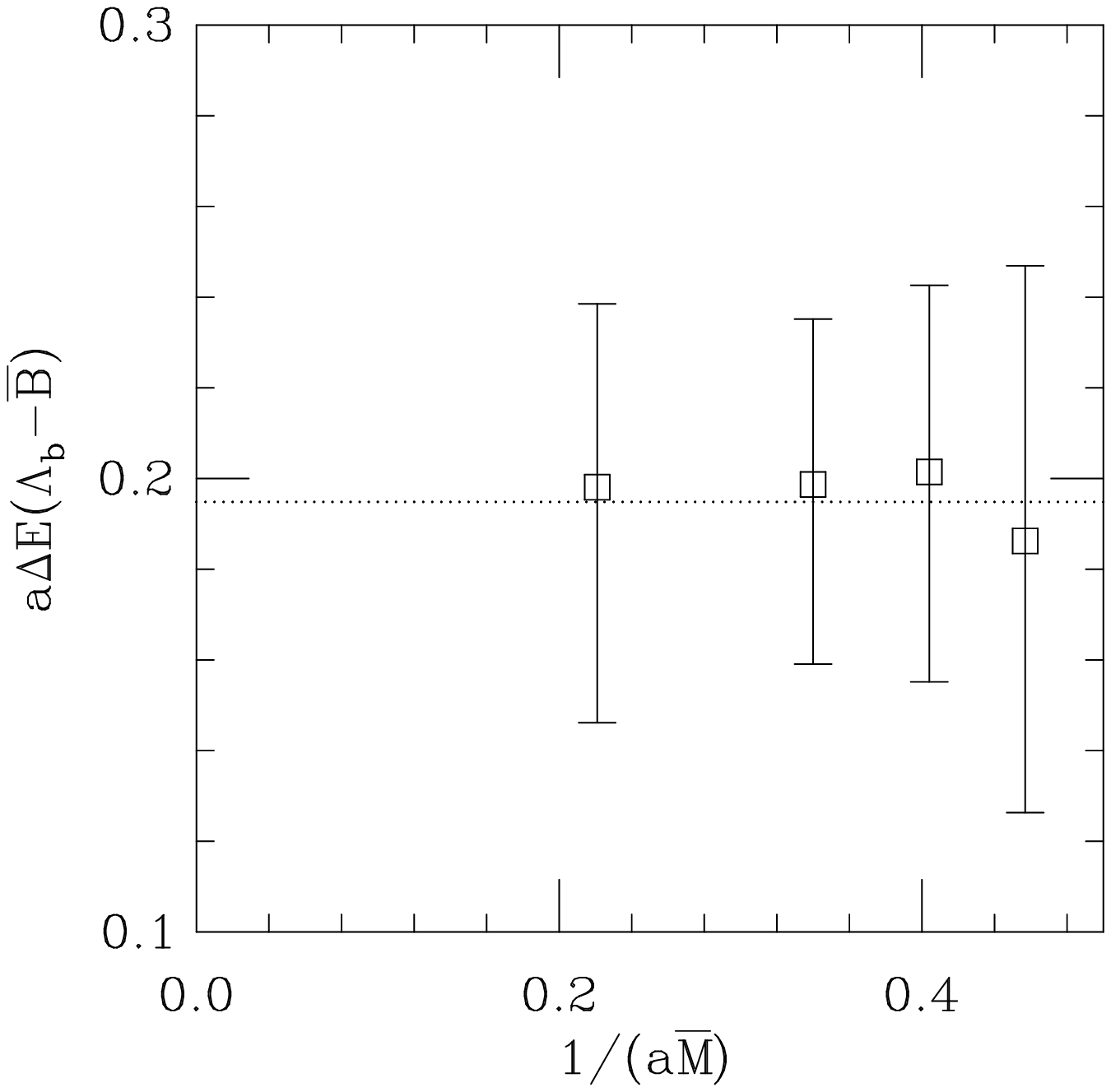}{13cm}}
  \caption{The chirally extrapolated splitting between the
$\Lambda_b$ and the
spin-average of the $B$ and $B^*$, plotted against the
inverse spin-averaged meson mass. The dotted line shows a correlated
fit of
the data to a constant.}
 \label{fig:Lambda_B_corr}
\end{figure}
An example for the
effective mass of the  correlation function of the $\Lambda_b$ baryon
is shown in figure~\ref{figure:effmass_Lb_B}. The bare binding energy
of the
$\Lambda_b$ is determined  from a fit of the correlation function to
a single
exponential. The interval is chosen to be
 $t_{min}/t_{max}$ = $7/13$ for $\kappa = 0.1440$ and
 $t_{min}/t_{max}$ = $7/16$ for $\kappa = 0.1432$. The results are
given in
table~\ref{table:fit_lambda_b_B}. The splitting between the
$\Lambda_b$ and
the meson is determined by calculating the bootstrapped difference
between the
ground state energies of both states.  On our ensemble there are
large error
bars on the baryon energy.
{}From HQET we expect that the splitting
between the $\Lambda_b$ and the spin average
of the $B$ and the $B^*$,  $\overline{B}$,
should be fairly independent  of the heavy quark mass. In the
$\Lambda_b$ the
$b$ quark couples to a spin zero system of light quarks. In the $B$
and the $B^*$, $b$  couples to a light quark of spin $1/2$, giving a
hyperfine
splitting which is a $1/m_Q$
effect. This expectation is fulfilled  well
for our results, as is shown in figure~\ref{fig:Lambda_B_corr}, and
is
borne out by experiment, $\Delta E(\Lambda_b-\overline{B})$ being
$330(50)$ MeV and
$\Delta E(\Lambda_c-\overline{D})$ being  $313(1)$ MeV. Performing a
correlated fit of the splittings at all our heavy quark masses to a
constant
$C$, one obtains $C = 0.20(3)$, with a $\chi^2$ per degree of freedom
of
0.01. The mass independence of this splitting makes it a good
quantity
to calculate and we could actually use it to extract $a^{-1}$.  Using
the
experimental value for the splitting at the $B$, we obtain $a^{-1} =
1.7(2)$
GeV.
Converting the constant $C$ into physical units, using a lattice
spacing of
$a^{-1}=2.05(10)$ GeV, one obtains 0.41(9) GeV.
Our central value is slightly higher  than the
experimental result, albeit with large errors.  It is possible that
the
$\Lambda_b$ suffers finite volume effects on these lattices.

$C$ is also  our prediction
for the static $\Lambda_b-B$ splitting,
 which is in  good agreement with
the static result of the UKQCD collaboration at $\beta = 6.2$ of
$0.42\errr{10}{9}(stat) \errr{3}{3}(syst)$ GeV~\cite{bbar}. Our
correlation
functions for the static
$\Lambda_b$ are rather noisy at this level of statistics and it is
problematic to extract baryon energies from them.

Other groups extract results for the $\Lambda_b-B$ splitting. This is
expected
to be more  dependent on the heavy quark mass. Experimentally, one
has
$\Delta E(\Lambda_b-B) = 0.36(5)$ GeV and
$\Delta E(\Lambda_c-D) = 0.416(1)$ GeV.
We work at the $b$ quark mass and can extract a splitting of
$0.43(10)$ GeV.
In fact there is no noticeable mass dependence in our $\Lambda_b-B$
splitting
in  table~\ref{table:res_lambda_b_B}, but our error bars are
larger than the hyperfine splitting, so at our level of statistics we
could
not  clearly detect it.
However, without chiral extrapolation
the mass dependence is somewhat more obvious. Other groups have used
a
relativistic  quark action and extrapolated from moderate
quark masses to the $b$ (for a compilation of results
see~\cite{nico}).
This is dangerous, particularly if such an extrapolation has been
shown to
give an incorrect hyperfine splitting. We believe that there could be
significant systematic errors attached to this procedure.
\subsubsection{$P-S$ splitting}
The observation of orbitally excited mesons, generically called
 $B^{**}$ states, has been
reported recently by DELPHI and OPAL~\cite{DELPHI,OPAL}. They study
decay
modes to $B\pi$ and $B^*\pi$ and are not able to clearly separate the
expected $0^+$, $1^+$, $1^{+'}$ and $2^+$ resonances, the first two
expected
to be rather broad ($j_q=1/2$) and the latter two, narrow
($j_q=3/2$).
A mass of 5.73(25) GeV is
quoted for the cross-section weighted mean mass of $B^{**}$ states by
DELPHI.
This gives some ambiguity when we wish to compare lattice
calculations
with experiment. The ideal quantity to calculate is the spin-averaged
splitting, $\overline{B^{**}} - \overline{B}$. The experiments
above give a $B^{**} - \overline{B}$ splitting of 419(25) MeV without
spin-averaging. Thus individual $B^{**}$ states, and hence the
spin-average
may differ from this value by the  splittings
between $P$ states, and it is unknown  how individual states
contribute to the
experimental number.
A splitting of around 50 MeV between the  $j_q=3/2$ and the $j_q =
1/2$
states has been suggested~\cite{rosner}. For the case
of charm the two narrow $2^{+}$ and $1^{+'}$ states have been clearly
seen
and yield a $D(1^{+'}) - \overline{D}$ splitting of 450(3) MeV. This
is very
similar to the number above and indicates that the heavy quark mass
dependence of the $P - \overline{S}$ splitting isn't large and will
only be visible once the separation of $P$ states is clear.
\begin{figure}[thbp]
\vspace{-2cm}
\centerline{\epsfxsize=10cm
\epsfbox{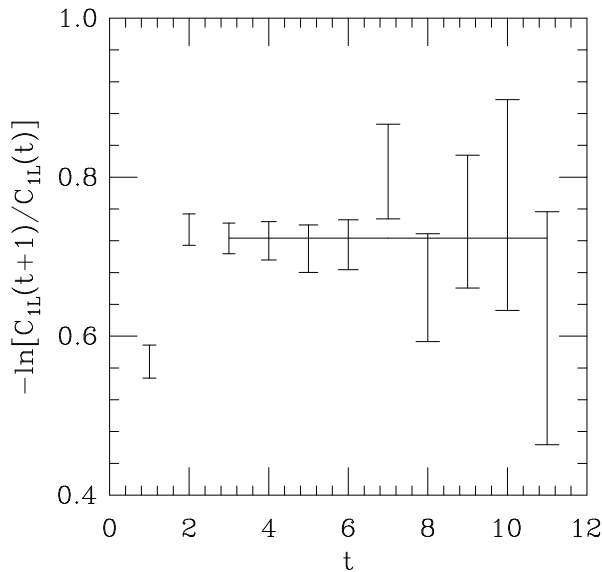}}
  \caption{Effective mass of the $^1P_1$ correlation function,
$am_Q^0$ = 2.0,
$\kappa = 0.1432$. The solid line denotes the fit. }
 \label{fig:effmass_1p1}
\end{figure}
\begin{table}
\begin{center}
\begin{tabular}{|c|c|c|c|c|}
\hline
                   \multicolumn{1}{|c|}{}
                 & \multicolumn{2}{c|}{$\kappa=0.1432$}
                 & \multicolumn{2}{c|}{$\kappa=0.1440$} \\
\hline
$am_Q^0$&$\;\;E_{\rm sim} \;\;$&Q&$\;\;E_{\rm sim} \;\;$&Q\\
\cline{1-5}
1.71  & 0.70(4) &0.39    & 0.72(4) &0.39  \\
2.0   & 0.72(1) &0.27   & 0.72(1) & 0.25\\
2.5   & 0.71(1) &0.26  & 0.71(1) & 0.23 \\
4.0   & 0.71(1) &0.32   & 0.71(2) & 0.38 \\
\cline{1-5}
\end{tabular}
\end{center}
\caption{Fit results for the $^1P_1$ states.}
\label{table:p_fit}
\end{table}

For our $P$ state correlation functions, we are only able to extract
a good signal from the $^1P_1$. The $^3P_0$, $^3P_1$ and $^3P_2$ are
rather
noisier and give, where visible, a very similar effective mass to
that of the
the $^1P_1$. We have not considered the $^1P_1/^3P_1$
cross-correlations,
 so it is not clear whether the mass we are extracting from the
$^1P_1$ is
that of the physical $1^{+}$ or the $1^{+'}$ or some average.
 This theoretical uncertainty,
which can be improved with a higher statistics calculation and taking
the
cross-correlations into consideration~\cite{B_c}, is
similar to the current experimental uncertainty described above.

For the heavy quark masses $2.0$, $2.5$ and $4.0$,
a fit interval of $t_{min}/t_{max} = 3/11$ was chosen for our $^1P_1$
correlation functions.  An example for
an effective mass plot is shown in figure~\ref{fig:effmass_1p1}. In
the run
with $am_Q^0 = 1.71$ the smearing functions for the $P$ states were
less
optimized, so there a fit interval of  $t_{min}/t_{max} = 8/12$ was
used.
The results for the $^1P_1$ seem to be  independent of the light
quark mass, so we choose to determine
the $S-P$ splitting by calculating the bootstrapped
difference between the $^1P_1$ simulation energy at $\kappa = 0.1432$
and
the chirally extrapolated spin-averaged $S$ state simulation energy.
\begin{table}
\begin{center}
\begin{tabular}{|c|c|}
\hline
 $am_Q^0$&$\;\;a \Delta E \;\;$\\
\cline{1-2}
$1.71$ & 0.27(4)  \\
$2.0$  & 0.27(2)  \\
$2.5$  &  0.25(2)  \\
$4.0$  &  0.25(2) \\
\hline
\end{tabular}
\end{center}
\caption{$^1P_1-\overline{S}$ splitting.}
\label{table:S-P}
\end{table}

\begin{figure}[hbtp]
\vspace{-3cm}
\centerline{\ewxy{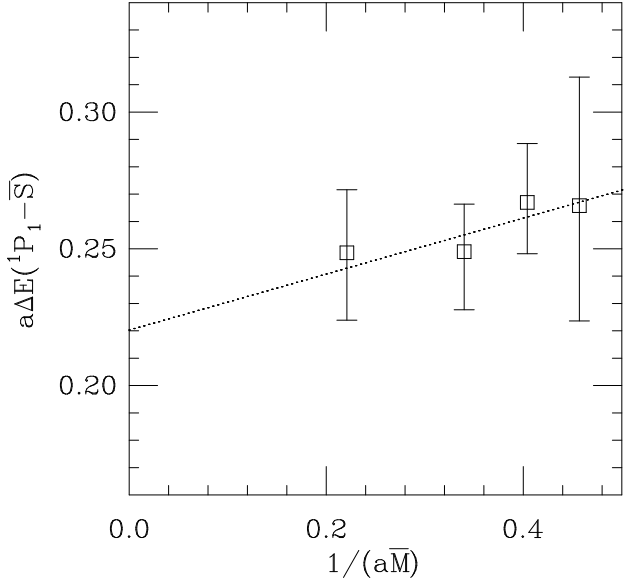}{13cm}}
  \caption{The $^1P_1-\overline{S}$ splitting plotted
against the
inverse spin-averaged meson mass. The dotted line corresponds to
 a correlated fit of
the lattice results to a linear  function.}
 \label{fig:S-P_corr}
\end{figure}

In figure~\ref{fig:S-P_corr} we show the $S-P$ splitting plotted
against the
inverse spin-averaged meson mass. The dependence on heavy
quark mass is clearly small, as expected. A correlated fit to a
linear function
gives an extrapolation to the static value of $0.22(5)$ in lattice
units and
a slope of $0.1(1)$, compatible with zero.
Taking our result at $am_Q^0 = 2.0$, which is approximately at the
$B$,
and $a^{-1}$ = 2.05(10) GeV gives a
splitting of $0.55(4)$ GeV. This is rather higher than the
experimental value above but not inconsistent given the systematic
errors arising from the uncertainty (both theoretical and
experimental) over which $B^{**}$ state we are considering.
If we use the experimental result to derive a lattice spacing,
we obtain the rather low value 1.6(2) GeV, in fact in
agreement with our result from $\Lambda_b - \overline{B}$.
Previous results in the static approximation \cite{eichtenp} give
a splitting of around 0.4 GeV for the $B(1^{+}) - \overline{B}$
splitting. There is some indication in our results that the static
value might be slightly lower than that at the $b$ mass, but we
have not attempted to calculate this splitting in the static
approximation.
\section{\label{sec:conclusions} Conclusions}
We present the first comprehensive lattice study of the heavy-light
spectrum
at  meson masses in the region of the $B$ in the quenched
approximation, using
NRQCD and static heavy quarks. The light fermion action has been
improved through $O(a)$ at tree level, and the NRQCD action has  been
tadpole improved. Work is in progress using tadpole improved light
fermions
with higher statistics~\cite{us_in_progress}.
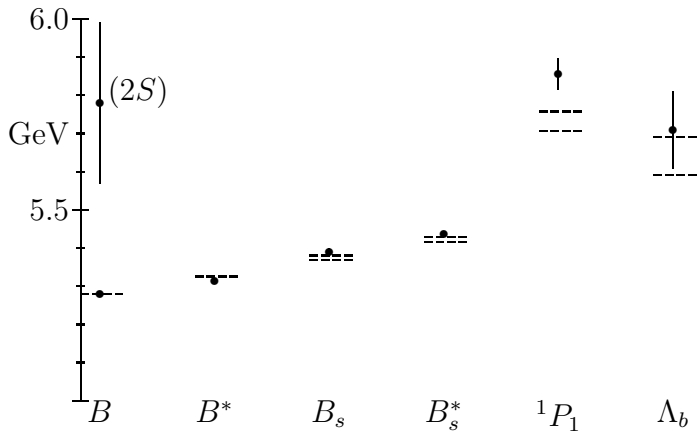
\begin{figure}
\begin{center}
\setlength{\unitlength}{.02in}
\begin{picture}(130,100)(50,500)
\put(15,500){\line(0,1){100}}
\multiput(13,500)(0,50){3}{\line(1,0){4}}
\multiput(14,500)(0,10){11}{\line(1,0){2}}
\put(12,550){\makebox(0,0)[r]{5.5}}
\put(12,600){\makebox(0,0)[r]{6.0}}
\put(12,570){\makebox(0,0)[r]{GeV}}

     \put(20,500){\makebox(0,0)[t]{$B$}}
     \put(20,528){\circle*{2}}
     \multiput(15,527.9)(3,0){4}{\line(1,0){2}}
     \put(20,578){\circle*{2}}
     \put(20,578){\line(0,1){21}}
     \put(20,578){\line(0,-1){21}}
     \put(30,585){\makebox(0,0)[t]{$(2S)$}}

     \put(50,500){\makebox(0,0)[t]{$B^{*}$}}
     \put(50,531.5){\circle*{2}}
     \put(50,531.5){\line(0,1){0.5}}
     \put(50,531.5){\line(0,-1){0.5}}
     \multiput(45,532.6)(3,0){4}{\line(1,0){2}}
     \multiput(45,532.4)(3,0){4}{\line(1,0){2}}

     \put(80,500){\makebox(0,0)[t]{$B_s$}}
     \put(80,539){\circle*{2}}
     \put(80,539){\line(0,1){0.8}}
     \put(80,539){\line(0,-1){1}}
     \multiput(75,538.1)(3,0){4}{\line(1,0){2}}
     \multiput(75,536.9)(3,0){4}{\line(1,0){2}}

     \put(110,500){\makebox(0,0)[t]{$B^{*}_s$}}
     \put(110,543.7){\circle*{2}}
     \put(110,543.7){\line(0,1){0.5}}
     \put(110,543.7){\line(0,-1){0.5}}
     \multiput(105,542.8)(3,0){4}{\line(1,0){2}}
     \multiput(105,541.6)(3,0){4}{\line(1,0){2}}

     \put(140,500){\makebox(0,0)[t]{$^1P_1$}}
     \put(140,585.6){\circle*{2}}
     \put(140,585.6){\line(0,1){4}}
     \put(140,585.6){\line(0,-1){4}}
     \multiput(135,570.7)(3,0){4}{\line(1,0){2}}
     \multiput(135,575.7)(3,0){4}{\line(1,0){2}}

     \put(170,500){\makebox(0,0)[t]{$\Lambda_b$}}
     \put(170,570.9){\circle*{2}}
     \put(170,570.9){\line(0,1){10}}
     \put(170,570.9){\line(0,-1){10}}
     \multiput(165,559.1)(3,0){4}{\line(1,0){2}}
     \multiput(165,569.0)(3,0){4}{\line(1,0){2}}

\end{picture}
\end{center}
  \caption{The $B$ spectrum. Filled circles denote our results, where
error bars do not take uncertainties in $a^{-1}$ and the perturbative
energy
shifts into account. The dashed lines
denote the upper and lower bounds on the experimental data. The mass
of the
$B$ has been shifted upwards to match the physical value. }
 \label{fig:spec}
\end{figure}

Our work has shown that with our method it is feasible to simulate
heavy-light
hadrons with a $b$ quark
directly on the lattice. Results for meson masses, the hyperfine
splitting,
$B_s-B_d$  splitting, $P$ wave states and heavy-light baryons have
been
obtained that are in reasonable agreement with experiment. A
comparison between
the lattice results and experimental values is shown in
figure~\ref{fig:spec}.
Note that in figure~\ref{fig:spec}
we have fixed the $B$ mass to its experimental value and
plotted
splittings in physical units.
This study on quenched configurations has been performed in parallel
with
an investigation with 2 flavours of dynamical
quarks~\cite{spec_dynamical,sara_bielefeld}, which enables us in
principle to
estimate the dependence of the results on the number of flavours.
\section*{Acknowledgements} We are grateful to the UKQCD
collaboration for
allowing  us to use their gauge configurations and light propagators.
We would like to thank C. Morningstar and G. P. Lepage for useful
discussions. This work was supported by SHEFC, PPARC and the U.S.
DOE. We
acknowledge support by the NATO under grant  CRG 941259 and the EU
under
contract CHRX-CT92-0051. We thank the Edinburgh Parallel Computing
Centre
for providing computer time on their CM-200.

\end{document}